\documentclass[useAMS, usenatbib, referee]{draftstyle}
\usepackage{graphicx}
\usepackage{float}
\usepackage{caption, setspace}
\usepackage{mathtools}
\def\bSig\mathbf{\Sigma}

\newcommand{\vect}[1]{\bmath{#1}}

\title[Estimating the Entropy Rate of Finite Markov Chains]{Estimating the Entropy Rate of Finite Markov Chains with Application to Behavior Studies}

\author
{Brian Vegetabile$^{1,*}$\email{bvegetab@uci.edu} 
and Jenny Molet$^{2}$
and Tallie Z. Baram$^{2,3,4}$
and Hal Stern$^{1}$ \\
$^{1}$Department of Statistics, University of California, Irvine, CA, U.S.A. \\
$^{2}$Department of Anatomy and Neurobiology, University of California, Irvine, CA, U.S.A. \\
$^{3}$Department of Pediatrics,  University of California, Irvine, CA, U.S.A. \\
$^{4}$Department of Neurology,  University of California, Irvine, CA, U.S.A.}

\begin{document}





\pagerange{\pageref{firstpage}--\pageref{lastpage}} 
\volume{XXX}
\pubyear{2017}
\artmonth{November}


\doi{xx.xxxx/x.xxxx-xxxx.xxxx.xxxxx.x}


\label{firstpage}


\begin{abstract}
Predictability of behavior has emerged an an important characteristic in many fields including biology, medicine, and marketing.  Behavior can be recorded as a sequence of actions performed by an individual over a given time period. This sequence of actions can often be modeled as a stationary time-homogeneous Markov chain and the predictability of the individual's behavior can be quantified by the entropy rate of the process.  This paper provides a comprehensive investigation of three estimators of the entropy rate of finite Markov processes and a bootstrap procedure for providing standard errors.  The first two methods directly estimate the entropy rate through estimates of the transition matrix and stationary distribution of the process; the methods differ in the technique used to estimate the stationary distribution.  The third method is related to the sliding-window Lempel-Ziv (SWLZ) compression algorithm.  The first two methods achieve consistent estimates of the true entropy rate for reasonably short observed sequences, but are limited by requiring a priori specification of the order of the process.  The method based on the SWLZ algorithm does not require specifying the order of the process and is optimal in the limit of an infinite sequence, but is biased for short sequences.  When used together, the methods can provide a clear picture of the entropy rate of an individual's behavior.
\end{abstract}

%



\maketitle


%

\section{Introduction}
\label{s:intro}

The behavior of a biological system is often characterized by recording a series of observed actions of one or more elements of the system.   Behavior can be summarized in terms of the actions that occur most frequently, the proportion of time that specific actions occur, or the variability in the types of actions performed.  Recently, \citet{molet2016fragmentation}, demonstrated that the patterns and rhythms in the observed behaviors of rodent mothers can have lasting consequences for their offspring.  Their study modeled rodent behavior as a discrete-state stochastic process, specifically a Markov chain, and quantified predictability of maternal behavior through the \textit{entropy rate} of the process, a measure of the predictability of a sequence of actions.  This paper compares approaches for estimating the entropy rate of an observed sequence and focuses on understanding the properties of these estimators for short sequences which are common in behavior studies.  
  
%

Entropy is a measurement that defines the predictability of a single random variable.  Entropy rate extends the concept of entropy from random variables to stochastic processes.   If we consider a sequence of random variables, entropy rate quantifies the limiting behavior of the joint entropy of the random variables as the sequence length increases.  For a stationary Markov chain, the entropy rate is a function of the stationary distribution and the transition matrix that defines the dependence structure of the process.  Thus for a finite Markov chain, one method of estimating the entropy rate is to estimate both the transition matrix that defines the behavior of the system and the limiting behavior of the system.  Another approach to estimating the entropy rate is based on Lempel-Ziv compression algorithms \citep{ziv1977universal}.  \citet{shannonmathematical} described a key relationship between the compressibility of a sequence and the joint entropy of the sequence.  In theory, Lempel-Ziv compression algorithms are optimal in achieving the compression limit put forth by Shannon for Markov processes of any order \citep{coverthomas1991, wyner1994sliding} and therefore can be used to estimate the entropy rate.

The entropy rate has been used in a wide range of applications to characterize the behavior of signals in the presence of noise.  Specifically it has been used in measuring of the predictability of human mobility \citep{Song1018, McInerney2013808}, assessing the complexity of short heart period variability series \citep{portaEntropyRate}, characterizing neural spike trains \citep{amigospikes}, classifying differences in behavior on Twitter \citep{chu2010tweeting}, quantifying the difference in behavioral patterns induced by distinct environments \citep{molet2016fragmentation}, and associating predictability of maternal behavior with emotional and cognitive outcomes of their infants and children \citep{davis2017exposure}.   \citet{gao2008estimating} provided a thorough analysis of entropy rate estimators for binary sequences and assessed performance of these estimators for long sequences.  This article reports on a comprehensive investigation of various entropy rate estimation techniques for finite state spaces and smaller sequence lengths typical in behavior studies.

In Section 2, we briefly review the definitions of Markov chains, entropy, and entropy rate.  Section 3 outlines techniques for estimating the entropy rate and a bootstrap procedure for obtaining the standard error of the entropy rate estimators.  Section 4 provides a simulation study comparing the performance of the estimators under varying sequence lengths and different data generating processes.   Section 5 applies the estimators in a biological study of patterns of maternal nurturing behaviors of rodent dams towards their pups.  Finally, Section 6 provides a summary of the methods and their properties.  

\section{Modeling}

\subsection{Modeling Behavior Using Finite-State Markov Chains}\label{sec:finitemarkov}

We model behavior as a sequence of observable actions through time.  Let $\mathcal{X} = \{X_t\}$ be a stochastic process composed of a sequence of random variables observed at time points, $t \in \mathcal{T} = \{0,1,\dots,T\}$. Further, we define the state space  $\mathcal{A} = \{ \alpha_1, \alpha_2, \dots, \alpha_\kappa \}$ to be the finite set of $\kappa$ allowable actions that an individual is capable of performing.  We observe $X_t = x_t$, with $x_t \in \mathcal{A}$  for all time points $t$.  Additionally, we assume that the probability distribution of the random variable $X_t$ depends only upon previously observed actions (i.e. the future does not affect the present). Then the joint distribution can be expressed as,  $\mbox{Pr}(X_0 = x_0, \dots, X_T = x_T) = \mbox{Pr}(X_0 = x_0) \cdot\prod_{t=1}^{T} \mbox{Pr}(X_t  = x_t | X_{t-1} = x_{t-1}, \dots, X_0 = x_0)$, where each $x_t \in \mathcal{A}$.  We simplify the dependence structure and assume that only the previous $m$ observations are relevant, then
\begin{eqnarray}
\mbox{Pr}(X_t = x_t | X_{t-1} = x_{t-1}, \dots, X_0=x_0) = \mbox{Pr}(X_t = x_t | X_{t-1} = x_{t-1}, \dots, X_{t-m}=x_{t-m}). 
\label{eqn:markovproperty}
\end{eqnarray}
A stochastic process with this assumption and a countable state space is known as an $m^{th}$-order Markov chain \citep[see][Chapter 1, Section 3c]{karlin1975first}.  

We briefly review required Markov chain theory using a first-order Markov chain (i.e. $m=1$) and then generalize below to higher order Markov chains.  Let $P_{ij} = P(\alpha_i, \alpha_j)= \mbox{Pr}(X_t = \alpha_j | X_{t-1} = \alpha_i)$ be the probability of transitioning from state $\alpha_i$ at time $t-1$ to state $\alpha_j$ at time $t$ (we assume the transition probabilities are independent of time so the process is time-homogeneous). Properties of the transition matrix $\bmath{P}$ are important for understanding the behavior of the process.    Let $\mu_{t}$ represent the marginal distribution of the random variable $X_{t}$. It is easy to show that the distribution at time $t+n$ is $\mu_{t+n} =  \mu_{t} \bmath{P}^n$. The matrix $\bmath{P}^n$ is referred to as the $n$-step transition matrix of the first-order Markov chain. 

In addition to assuming the order of the process is known and that the process is time-homogeneous, we also assume that the Markov chain is irreducible and stationary (we rely on results which apply to processes of unknown periodicity).  \citet{levin2009markov} provide that a Markov chain is \textbf{irreducible} if for any two actions $\alpha_i, \alpha_j \in \mathcal{A}$ there exists an integer $t$ (possibly depending on $\alpha_i$ and $\alpha_j$) such that $\bmath{P}^t(\alpha_i, \alpha_j) > 0$ and therefore there is no absorbing state in the observed process.  The assumption of stationarity implies that the joint distribution on sets of random variables $(X_{t_1+h}, X_{t_2+h}, \dots, X_{t_n+h})$ and  $(X_{t_1}, X_{t_2}, \dots, X_{t_n})$ are the same for all $h > 0$ and arbitrary $t_1, t_2, \dots, t_n$ from $\mathcal{T}$ \citep{karlin1975first}.  For a Markov process to be a stationary process $\mu_t$, the distribution of $X_t$,  and $\mu_{t-1}$, the distribution of $X_{t-1}$, must be the same.  This common distribution is known as the stationary distribution $\pi = (\pi(\alpha_1), \pi(\alpha_2), \dots, \pi(\alpha_k)) = \left( \pi_1, \pi_2, \dots, \pi_{\kappa}\right)$. Additionally, since $\mu_{t} = \mu_{t-1} \bmath{P}$ it follows that the stationary distribution of a Markov process must satisfy $\pi = \pi \bmath{P}$. The stationary distribution $\pi$ is not guaranteed to exist for all Markov chains, but the assumption of irreducibility implies that distribution $\pi$ will both exist and be unique \citep[see][Proposition 1.14 and Corollary 1.17]{levin2009markov}.   A useful interpretation of the stationary distribution is that it is the asymptotic proportion of time that the Markov chain will spend in any state \citep{levin2009markov}.  

Now we generalize these definitions to higher-order Markov processes.  To do this, we first develop a first-order Markov chain \textit{on vectors} of random variables.  Let $\mathcal{Y} = \{\bmath{Y_t}\}$ be a stochastic process composed of a set of random vectors with $m$ elements each and a discrete time indexing.   As an example that is relevant below, we may have, $\bmath{Y_t} = (Y_{t,0}, Y_{t,1}, \dots, Y_{t,m-1})^T$, a column vector of random variables, each with state space $\mathcal{A}$.  The state space of each $\bmath{Y_t} \in \mathcal{Y}$ is therefore the $m$-fold Cartesian product of the state spaces of the component random variables, denoted as $\mathcal{B}$,  and each $\beta_{\bmath{i}} \in \mathcal{B}$ represents an ordered $m$-tuple of states or actions.  We can construct transition matrices for the vector process, but now the matrix gives the probability of transitions between $m$-tuples.  Definitions of stationarity and irreducibility now follow as in the earlier discussions. 

To relate $m^{th}$-order Markov chains of random variables to first-order chains of random vectors, note that we can construct a vector process from the original $m^{th}$-order univariate process using contiguous subsequences of random variables of the original sequence.  We define $\bmath{X_i^m}$ as the subsequence of $\{X_t\}$ starting from index $i$ and composed of $m$ contiguous observations, or $\bmath{X_i^m} = (X_{i+(m-1)}, X_{i+(m-2)}, \dots, X_{i})^T$.  Now if we consider the probability of transitioning from $\bmath{X_{t-1}^m}$ to $\bmath{X_{t}^m}$, it follows that $\mbox{Pr}(\bmath{X_{t}^m} = \bmath{x_t} | \bmath{X_{t-1}^m} = \bmath{x_{t-1}}) = \mbox{Pr}(X_{t+m-1} = x_{t+m-1} | X_{t+m-2} = x_{t+m-2}, \dots, X_{t-1} = x_{t-1})$, where the expression follows because $\bmath{X_{t}^m}$ and $\bmath{X_{t-1}^m}$ share $m-1$ common random variables. This demonstrates that first-order transitions of vectors of the newly constructed process are equivalent to $m^{th}$ order transitions of the original process.  The advantage of this construction is that we can find an appropriate first-order transition matrix for the process of random vectors and utilize all of the definitions which we have previously outlined.  It is worth noting that under this construction the transition matrix will be of dimension $(\kappa^m \times \kappa^m)$ and may pose computational issues for large $m$ or if the cardinality of $\mathcal{A}$ is large.  

\subsection{Entropy Rate of a Finite Markov Chain}

\citet{shannonmathematical} introduced the concept of entropy in the context of communication.  The entropy of a discrete random variable $X$, taking values from the state space $\mathcal{A}$, is $H(X) = - \sum_{\alpha_i \in \mathcal{A}} \mbox{Pr}(X=\alpha_i) \log_2 \mbox{Pr}(X=\alpha_i).$  Entropy varies between zero and $\log_2 \kappa$, where $\kappa$ is the cardinality of $\mathcal{A}$.  The definition of entropy is easily generalized to joint distributions of random variables, as well to conditional distributions \citep[for an overview see][]{coverthomas1991}. 

\citet{coverthomas1991} define the entropy rate of a stochastic process $\mathcal{X} = \left\{ X_t\right\}$,  $t \in \{0, 1,\dots, T\}$ as, $H(\mathcal{X}) = \lim_{T\rightarrow\infty} \frac{1}{T} H(X_0, X_1, \dots, X_T)$ and further show that if the process is stationary this is equivalent to $H(\mathcal{X}) = \lim_{T\rightarrow\infty} H(X_T | X_{T-1}, X_{T-2}, \dots, X_0)$ provided that the limit exists.  The entropy rate of a stationary process provides a quantification of the predictability of the next observation given the history of observations which occurred before it.

We apply this definition to a stationary first-order time-homogenous Markov process, $\mathcal{X} = \left\{ X_t\right\}, t\in \{0,1,\dots, T\}$, with finite state space $\mathcal{A} = \{ \alpha_1, \alpha_2, \dots, \alpha_\kappa \}$.  Utilizing the framework outlined in Section \ref{sec:finitemarkov}, we find that the entropy rate of the process is\begin{eqnarray*}
H(\mathcal{X}) &=& \lim_{T\rightarrow\infty} H(X_T | X_{T-1}, X_{T-2}, \dots, X_0) \nonumber \\
&=& \lim_{T\rightarrow\infty} H(X_T | X_{T-1}) \nonumber \\
&=& \lim_{T\rightarrow\infty}  -\sum_{\alpha_{i} \in \mathcal{A}}\mbox{Pr}(X_{T-1} = \alpha_{i}) \sum_{ \alpha_{j} \in \mathcal{A}} \mbox{Pr}( X_{T} = \alpha_{j} | X_{T-1} = \alpha_{i}) \log_2 \mbox{Pr}( X_{T} = \alpha_{j} | X_{T-1} = \alpha_{i}) . \nonumber 
\end{eqnarray*}
The stationarity assumption and the assumption of a time-homogeneous Markov process implies $\mbox{Pr}(X_{T}=\alpha_{j} | X_{T-1} = \alpha_{i}) = P(\alpha_{i}, \alpha_{j}) = P_{ij}$. Noting that $\lim_{T\rightarrow\infty} \mbox{Pr}(X_{T-1} = \alpha_{i}) = \pi_{i}$ for all $\alpha_{i} \in \mathcal{A}$, we can write $H(\mathcal{X}) =  -\sum_{\alpha_{i}, \alpha_{j}\in \mathcal{A}} \pi_{i} P_{ij} \log_2 P_{ij}$.  The entropy rate can be seen to be a weighted average of the conditional entropy of  $X_{T}$ given the previous state $X_{T-1} = \alpha_{i}$, where the weights are given by the stationary distribution.  These results generalize easily for $m>1$ using the approach laid out in Section \ref{sec:finitemarkov}.

\subsection{Entropy Rate from Lempel-Ziv Compression} \label{sec:LZ}

Several authors \citep{amigospikes, Song1018, McInerney2013808} have utilized estimators derived from properties of Lempel-Ziv compression algorithms to estimate the entropy rate.  We briefly describe the idea behind Lempel-Ziv compression and its relationship to entropy.

\citet{lempel1976complexity} introduced methods  to quantify the complexity of finite sequences, as well as a framework for parsing sequences for compression.  They were able to relate their measure of the complexity of a sequence to the entropy rate of the source which generated that sequence.  They further provided algorithms \citep{ziv1977universal, ziv1978compression}  to provide compression of the sequence that can also utilized for estimating the entropy rate of the source.  We use the algorithm developed in 1977, which we will refer to as the sliding-window Lempel-Ziv, or SWLZ, algorithm. The optimality of the SWLZ algorithm for achieving a compression ratio approaching the entropy of a stationary ergodic source was provided in \citet{wyner1989some}, and investigated further in \citet{wyner1994sliding} and  \citet{ornstein1993entropy}.

To understand the SWLZ algorithm, consider a sequence of $n$ observations, $x_0 x_1\dots x_n$, as a realization from a stochastic process $\mathcal{X}$, with finite state space $\mathcal{A}$.  Further define $x_{i}^{j}$ to be the subsequence $x_i x_{i+1} \dots x_{j-1} x_j$, so that $x_{0}^{i-1}$ is the history of the subsequence before observation $x_i$.  The objective of the algorithm is to sequentially process the observations from $x_0$ to $x_n$ and partition the original sequence into unique subsequences.  To create these unique subsequences, consider that at some point $i \in \{1,\dots, n\}$ the previous $i-1$ observations have been parsed into subsequences.  The algorithm identifies the shortest length $L$ such that $x_{i}^{i+L-1} \not \subseteq x_{0}^{i-1}$, i.e. such that the sequence of length $L$ starting at $x_i$ has not been observed before.  At this point the substring $x_{i}^{i+L-1}$ is unique and we add it as a new parsing and move to position $x_{i+L}$.  This continues until the entire sequence is parsed, noting that the last subsequence may not be unique when all characters are exhausted.  Table \ref{tab:lz77} shows a unique parsing of a three state stochastic process.
\begin{table}
\centering
\caption{Example parsing based upon the SWLZ algorithm.}
\begin{tabular}{r|c}
  \hline
\texttt{Original Sequence} & 13131213232331313332 \\
\hline
\hline
\texttt{SWLZ Parsing} & $1 | 3 | 131 | 2 | 132 | 323 | 31313 | 332$ \\
   \hline
\end{tabular}
\bigskip
\label{tab:lz77}
\end{table}
In this example, note that the first symbol is unique $(\{1\})$ and that the second symbol is not equal to the first $(\{3\})$, so they both are unique subsequences.  When we start from the third observation we see that that $x_2 = 1$ is contained in the history $x_0 x_1 = 13$ and so is $x_2 x_3 = 13$, but the sequence $ x_2 x_3 x_4 = 131$ is a unique subsequence which has not been seen before. The algorithm stores this new subsequence and the process is continued then from observation $x_5$.  

The primary theoretical argument of the optimality of this algorithm relies on considering a \textit{doubly-infinite} sequence, $x_{-\infty}\dots x_{-2}x_{-1}x_0 x_1\dots x_{\infty}$ from a stationary ergodic process $\mathcal{X}$ with $x_j \in \mathcal{A}$. For $n=1,2,\dots$, \citet{wyner1989some} define $L_n(\mathcal{X})$ to be the smallest integer $L>0$ such that $x_{0}^{L-1}$ does not appear as a contiguous subsequence of $x_{-n}^{-1}$, or $L_n(\mathcal{X}) = \arg \min_{L} x_{0}^{L-1} \not \subseteq x_{-n}^{-1}$.
Note that $L_{n}(\mathcal{X})$ is the length of the unique parsing using a history of size $n$ by our previous definitions.  Under the assumption of a Markov process, \citet{wyner1989some} show that the entropy rate of the source $\mathcal{X}$ is related to the length of this parsing size and the size of the history $n$ used to create that parsing,
\begin{equation}
\frac{\log_2(n)}{L_{n}(\mathcal{X})} \rightarrow_{P} H(\mathcal{X}), \mbox{ as } n \rightarrow \infty.
\label{eqn:lztheory}
\end{equation}
These theoretical results are used below as the basis for an estimator of the entropy rate.

\section{Estimating the Entropy Rate}\label{sec:methods}

\subsection{Direct Estimation of Entropy Rate}

One approach to estimating the entropy rate is to estimate the transition matrix and stationary distribution and then apply the formulas of the previous section.  Directly estimating the transition matrix $\bmath{P}$ is straightforward using the observed transitions.  To estimate $\pi$, we describe two methods: 1) estimation of the stationary distribution based upon the observed proportion of time the process visits each state;  2) estimation of the stationary distribution based upon an eigenvalue decomposition of the transpose of the  estimated transition matrix. Once we have estimated $\bmath{P}$ by $\bmath{\hat P}$ and then $\pi$ by $\hat \pi$, we can estimate the entropy rate $H(\mathcal{X})$ as follows $\hat H(\mathcal{X}) =  -\sum_{i} \sum_{j} \hat\pi_{i} \hat P_{ij} \log_2 \hat P_{ij}$.

\subsubsection{Estimation of Transition Probabilities}

The transition matrix $\bmath{P}$ can be estimated from the observed transitions of the Markov chain.  If we denote $n_{ij}$ as the total number of transitions from $i^{th}$ row action to the $j^{th}$ column action, we can create a matrix $\bmath{T}_C = \{n_{ij}\}$ of transition counts.  
To convert this to a valid estimator of the transition matrix, we normalize each row by its corresponding row total.  Define $n_{i+}$ to be the sum of the counts over all columns $j$ in row $i$, $n_{i+} = \sum_{j} n_{ij}$.   The empirical estimator for the transition matrix, $\bmath{\hat P} = \left\{\frac{n_{ij}}{n_{i+}}\right\}$, is the maximum likelihood estimator for $\bmath{P}$ \citep[Section 17.2]{murphy2012machine}.

\subsubsection{Estimation of the Stationary Distribution}

~\\\textit{Using Observed Frequencies of States.} The stationary distribution of the Markov process can be estimated by considering the proportion of time each action is observed in a realization of this process.  Thus an empirical estimator of $\pi$ would be, $\hat \pi_{emp} = \left\{\frac{n_{i+}}{ n_{++}} \right\}$, where $n_{i+}$ is as defined previously and $n_{++} = \sum_{i} n_{i+} = \sum_{i} \sum_{j} n_{ij}$ is the total number of transitions. It can be shown algebraically that $\hat \pi_{emp} \ne \hat \pi_{emp} \bmath{\hat P}$, but the discrepancy will be small if the number of observations of the process is large.  

\textit{Using an Eigendecomposition of $\bmath{\hat P}^{T}$}\label{sec:eigenest}. An alternative is to estimate $\pi$ from an eigendecomposition of $\bmath{\hat P}^{T}$. For a stationary Markov process the stationary distribution is a row vector and must satisfy $\pi = \pi \bmath{P}$, or equivalently, $\pi^T = \bmath{P}^T \pi^T$, which means that the stationary distribution, $\pi$, is the transpose of the eigenvector corresponding to an eigenvalue equal to $1$ of $\bmath{P}^T$.  Because we have assumed an irreducible process there will be such an eigenvector \citep[see][Chapter 12, Theorem 3.1]{karlin1981second}

An estimator for the stationary distribution can be obtained by performing an eigendecomposition of the matrix $\bmath{\hat P}^{T}$ and setting $\hat\pi_{eig}$ proportional to the relevant eigenvector (i.e., an eigenvector that corresponds to the eigenvalue of $\lambda=1$).  Define $\hat x_{\lambda=1}$ to be this eigenvector and obtain an estimate of the stationary distribution as follows, $\hat \pi_{eig} = \hat x_{\lambda=1}^T/ \sum_{i=1}^{k}\hat x_{i, \lambda=1}$, where the denominator ensures that this is a valid probability distribution. 

\textit{Using a Limit of $\bmath{\hat P}^n$}.  There is another way to estimate $\pi$ by taking the limit of an infinite number of ``transitions'' based upon an estimated transition matrix $\bmath{\hat P}$.   For an irreducible Markov chain, \citet{kemeny1960finite} provide Theorem 5.1.4 which states that $\bmath{P}^n$ will be Cesaro-summable to a matrix $\Pi$ where each row of $\Pi$ is the stationary distribution $\pi$, i.e. $\lim_{n\rightarrow\infty} \frac{1}{n} \sum_{i=1}^n P^i(\alpha_{j}, \cdot) = \pi \mbox{ for all } \alpha_{j} \in \mathcal{A}$. Therefore for a very large value of $N$, $\pi$ can be estimated as, $\hat \pi_{limit} = \frac{1}{N} \sum_{i=1}^N \hat P^i(\alpha_{\vect{1}}, \cdot)$. This estimator achieves very slow convergence to the true $\pi$.  Additionally, there are computational challenges to taking powers of matrices.  Due to these computational and convergence issues, we do not provide simulation results for this estimator. 
%
%
%

\subsection{Sliding Window Lempel-Ziv Entropy Rate Estimation}

An alternative to estimating the entropy rate is to rely on the theoretical results for the asymptotic optimality of the SWLZ compression algorithm for stationary ergodic sources. Recall that the sequence in Equation (\ref{eqn:lztheory}), $\log_2(n) / L_n(\mathcal{X})$, converges in probability to the entropy rate of a Markov process, where $L_n(\mathcal{X})$ is the length of a unique parsing of the observed sequence when using a history of size $n$.  \citet{kontoyiannis1998nonparametric} show that a Ces\`aro summation of this series will also converge if the source is a stationary ergodic Markov process and that an estimator based on this relationship will be asymptotically consistent for estimating the entropy rate. 

Here we consider a modified version of the SWLZ  algorithm that is convenient for finding the entropy rate of a Markov chain.  Defining $\Lambda_{i}(\mathcal{X})$ to be the length of the shortest substring $x_{i}^{i+L-1}$ that does not appear as a substring in the history of $i$ symbols, $x_{0}^{i-1}$, Theorem 1C of  \cite{kontoyiannis1998nonparametric} provides that if $\mathcal{X} = \{X_i\}, i \in Z$ is a two-sided stationary ergodic Markov process with entropy $H(\mathcal{X})>0$, then 
$\lim_{n\rightarrow\infty} \frac{1}{n} \sum_{i=1}^{n} (\Lambda_i(\mathcal{X})/ log_2 n) = (H(\mathcal{X}))^{-1}$, converges almost surely. This result becomes the basis for an estimator.   

To estimate $H(\mathcal{X})$, we consider a realization of the process $\mathcal{X}$, to be the observed sequence, $x_0x_1 \dots x_T$.  At each instance $i$ of the observed sequence, we compute $\Lambda_i(\mathcal{X}) = \arg \min_{L} x_{i}^{i+L-1} \not \subseteq x_{0}^{i-1}$.  We outline the process in Table \ref{tab:lzalg}. To estimate the entropy rate we divide the bits required to encode a sequence length of $n$, $\log_2n$, by the average size of a unique parsing from this process. That is:
\begin{equation}
\hat H(\mathcal{X}) = \left[ \frac{1}{n} \sum_{i = 1}^{n} \frac{\Lambda_i(\mathcal{X})}{log_2n} \right]^{-1} = \frac{\log_2n}{\frac{1}{n}\sum_{i=1}^n \Lambda_i(\mathcal{X})}
\label{eqn:lzestimator}
\end{equation}
This version of the estimator considers a window size, or history, which  ``expands'' at each point $i$.  There are alternative versions of the estimator which consider fixed window sizes.  It has been our experience that the expanding window estimators perform the best for short sequence lengths (see \citet{gao2008estimating} for a broader comparison of those methods).  

\begin{table}
\centering
\caption{Step $i$ of the Sliding Window Lempel Ziv Algortihm}
\begin{tabular}{c|c|l|c}
  \hline
  \multicolumn{4}{l|}{\textbf{Original String:} $x_0x_1\dots x_{i-1} x_i x_{i+1} \dots x_{T-1} x_T$}  \\
  \hline
  Step & History & Candidate String & $x_{i}^{i+j-1} \subseteq x_{0}^{i-1}$ \\
  \hline
  i.1 & $x_0x_1\dots x_{i-1}$ & $x_i$ & \texttt{TRUE}\\
  i.2 & $x_0x_1\dots x_{i-1}$ & $x_ix_{i+1}$ & \texttt{TRUE} \\
  \vdots & \vdots & \vdots  & \vdots \\
  i.(L-1) & $x_0x_1\dots x_{i-1}$ & $x_ix_{i+1}\dots x_{i+L-2}$ & \texttt{TRUE} \\
  i.L & $x_0x_1\dots x_{i-1}$ & $x_ix_{i+1}\dots x_{i+L-1}$ & \texttt{FALSE} \\
  \hline
  \multicolumn{3}{r|}{\textbf{Size of Unique Parsing, $\Lambda_i(\mathcal{X})$}}  &  $L$ \\
\end{tabular}
\bigskip
\label{tab:lzalg}
\end{table}

\subsection{Estimation of Standard Errors via the Stationary Bootstrap}\label{sec:bootstrap}

The previous sections provide methods for obtaining point estimates of the entropy rate of a finite Markov chain. In this section we provide a method to measure the standard error of these point estimates.   A common approach to measuring the standard error of an estimate is to use the bootstrap method \citep{efron1994bootstrap}.  This method works well when observations are independent and identically distributed, but our application is focused on dependent data.  We therefore use a method called the ``stationary bootstrap'' of \citet{politis1994bootstrap} which is a variant of the block bootstrap.

A common approach to creating bootstrap samples of dependent data is that of the block bootstrap \citep{efron1994bootstrap}. The simple block bootstrap begins by first partitioning a sequence of observations into blocks of equal size.  To create a new bootstrap sequence, blocks are sampled with replacement from the partition and concatenated to create a sequence of approximately the same length as the original sequence.  An estimate of the quantity of interest is calculated on this new sequence and this process is repeated many times.  The standard error of the estimate is measured using the distribution of the point estimates across the bootstrap replicates.  An alternative version of the block bootstrap considers blocks of equal size, but the blocks are allowed to overlap.  Both methods are attractive in that they keep dependence between observations in the data, but are complicated by the need to choose the block size.

The stationary bootstrap is an extension of the block bootstrap which uses variable block sizes.  We describe the algorithm laid out in \citet{politis1994bootstrap} in the context of our notation. First define $C(i, l) = x_{i}^{i+l-1}$, where if $j\in \{i, i+1, \dots, i+l -1\} \ge \mathcal{T} +1$ we set $j \equiv j \bmod{ \mathcal{T} + 1}$.  Additionally define sequences of independent random variables $\mathcal{I}_{1}, \mathcal{I}_{2}, \dots $ and $L_{1}, L_{2}, \dots $ such that $\mathcal{I}_{i}$ is distributed discrete uniform on the set $\{0,1,\dots, \mathcal{T}\}$ and $L_i \sim Geometric(p)$ and let the random subsequence, $X_{k}^{*}$ be defined as $X_{k}^{*} = C(\mathcal{I}_{k}, L_k)$.  Finally define $\mathcal{X}^{*} = X_0^{*}\dots X_{k-1}^{*}X_{k}^{*}$ as the concatenation of the $k+1$ random subsequences.  While $\sum_{i=0}^{k} L_{i} \le \mathcal{T}+1$, we draw $\mathcal{I}_{k+1}$ and $L_{k+1}$ and set $\mathcal{X}^{*} = X_0^{*}\dots X_{k}^{*}X_{k+1}^{*}$.  Once the length of $\mathcal{X}^{*}$ is greater than $\mathcal{T}+1$ we take the first $\mathcal{T}+1$ observations and define that to be the bootstrapped sample $\mathcal{X}^{b}$.   For each bootstrap replicate we estimate the entropy $H(\mathcal{X}^{b})$ and then estimate the standard error of the bootstrap replicates.  

An advantage of the method is that given the original sequence of observations, the new sequence is stationary \citep[see][Proposition 1]{politis1994bootstrap}, while the traditional block bootstraps are not.  What is difficult is that the method requires specifying the parameter $p$, which is analogous to choosing a block size in the traditional block bootstrap.  Fortunately, the SWLZ estimate suggests a natural approach.  Recall that our estimator of the entropy rate is $\log_{2}(n)$ divided by the average unique block length.  This suggests that the average unique block length is approximately equal to $\log_{2}(n) / \hat H(\mathcal{X})$.  The average block length in the block bootstrap is $E(L_i) = p^{-1}$.  Equating these two values suggests choosing $p$ equal to $ \hat H(\mathcal{X}) / \log_{2}(n)$.  Section \ref{sec:bootstrapsims} provides a simulation which demonstrates the performance of this method for selecting $p$.

\section{Simulations}

Our research is motivated by an application of Markov chains to the study of behavior.  Before exploring the application, we carry out a simulation study to illustrate the performance of the three entropy rate estimators outlined in Section \ref{sec:methods}.  We compare the methods when we have correctly specified the model (the order of the Markov process) and when the order of the Markov process is misspecified.  

\subsection{Estimation of First-Order Markov Processes}\label{sec:firstordersims}

\begin{figure}
\centering 
\includegraphics[width=0.32\linewidth]{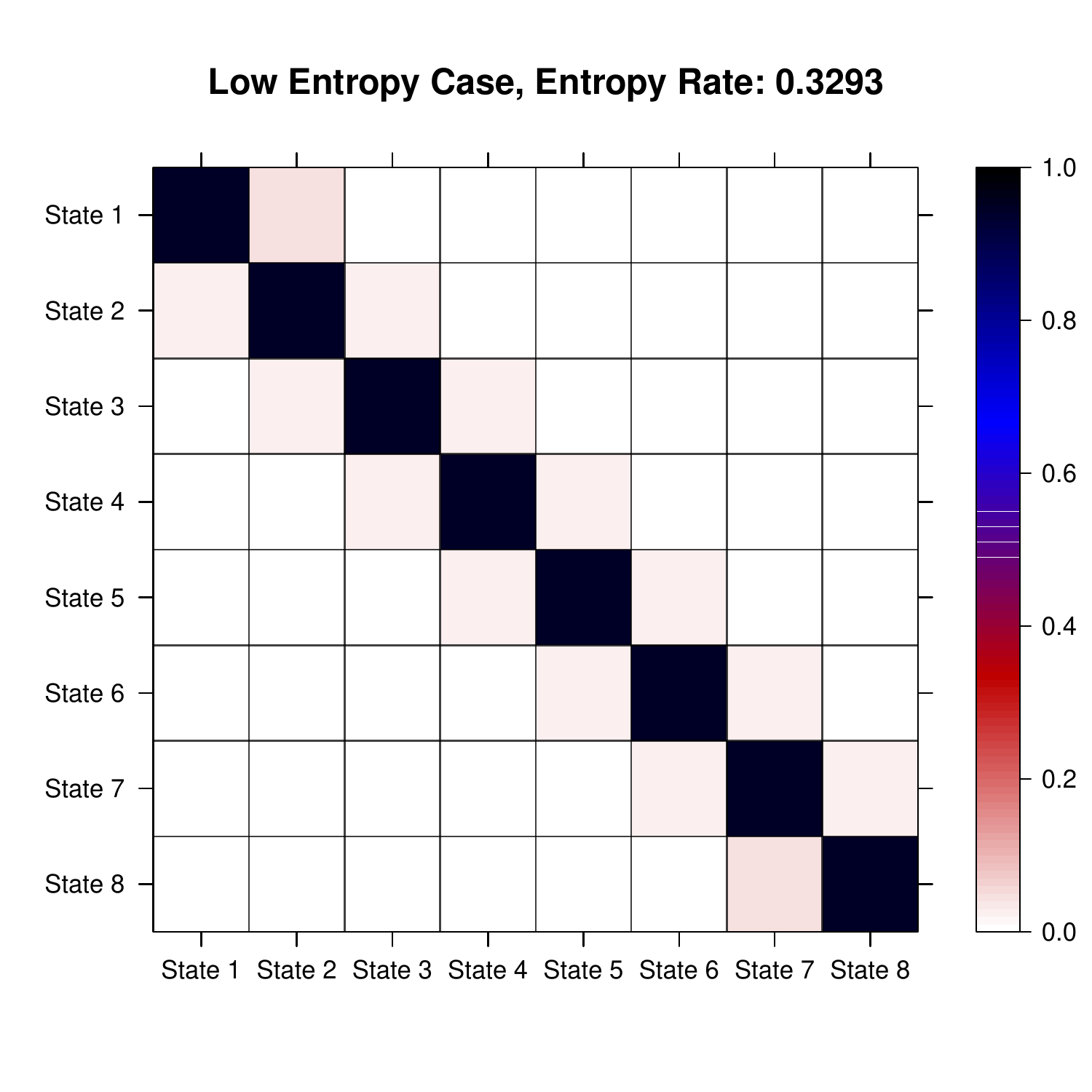}
\includegraphics[width=0.32\linewidth]{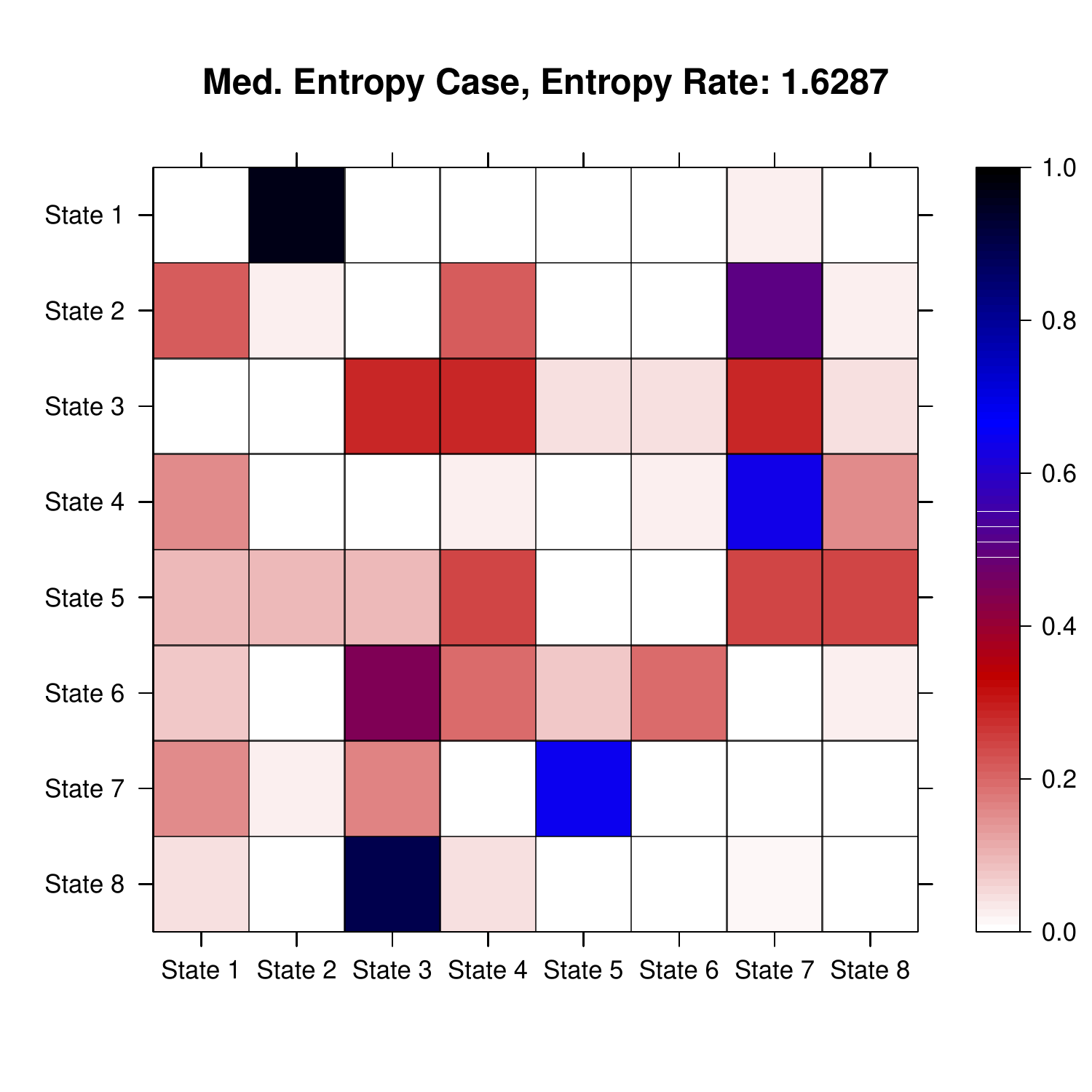}
\includegraphics[width=0.32\linewidth]{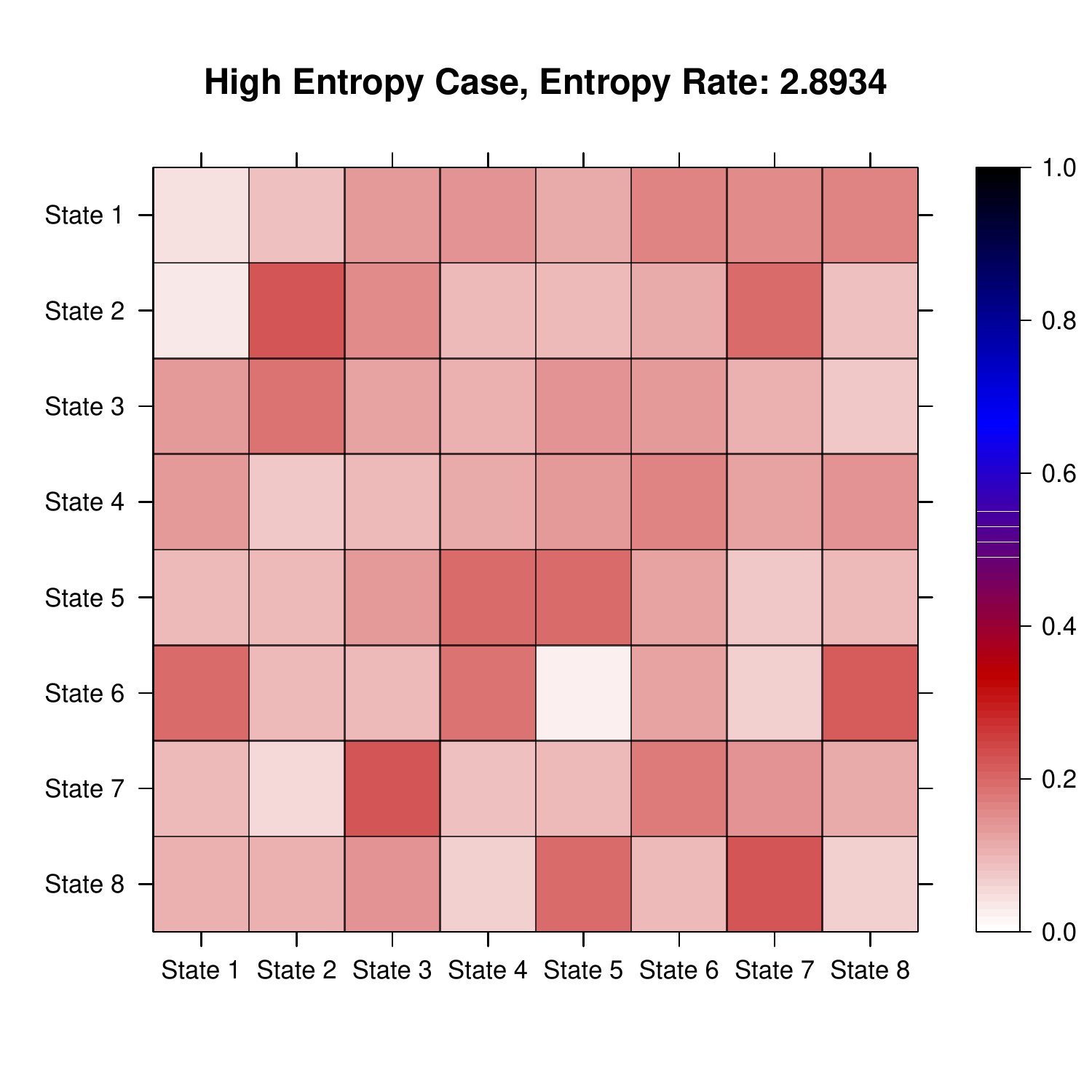}
\caption{Transition matrices describing three different types of behaviors: low entropy rate, intermediate entropy rate, and an instance of high entropy rate.}
\label{fig:transitionmatrices}
\end{figure}

The first setting of the simulation study is stationary time-homogeneous first-order Markov processes, with eight unique states.  We consider three different data generating models: a low entropy rate case, a medium entropy rate case, and a high entropy rate case. Figure \ref{fig:transitionmatrices} provides a visualization of the transition matrix for each data generating model with each box representing $P_{ij}$ (scale indicated at right).  The true entropy rate for each process is listed in the figure.  The entropy rate for a Markov process with eight states lies between 0 and 3.  

The low entropy rate transition matrix simulates a system where  there is a high probability of transitions to the same state ($P_{ii}=0.95$).  The high entropy rate transition matrix is a much less organized system, designed to behave similar to a purely random system $(P_{ij} \approx 0.125)$.  The medium entropy rate transition matrix is designed to be a balance between predictability and unpredictability, with the characteristic that for some states the next state is more predictable than others, and has an entropy rate that is similar to that found in our motivating example. 

We assess the performance of the various estimators as a function of observed sequence length. In a behavioral setting we are often interested in determining the number of observations required to obtain reliable inference and this simulation study will help inform this decision as well as provide information about the trade-offs between the estimation methods. Each Markov chain was simulated 100 times, with each simulation consisting of 10000 observations from the process.  We observe $\mathcal{X}_i = x_0x_1,\dots x_{9999}$ for $i =  1,\dots 100$.  We applied the estimation procedures of Section 3 to different length subsequences of each realization of the Markov chain and calculated an estimate of the entropy based on $x_{0}^{j-1}$ for $j \in \{50, 250, 500, 1000, 5000, 10000\}$.  The smaller values were chosen because behavioral applications, such as our motivating example, typically observe shorter sequence lengths of approximately 250 to 1000, while 5000 and 10000 were chosen to explore performance for longer sequence lengths.  \citet{gao2008estimating} provides a discuss for entropy rate estimation for sequences of much longer length. 

\begin{figure}
\centering
\includegraphics[width=0.9\linewidth]{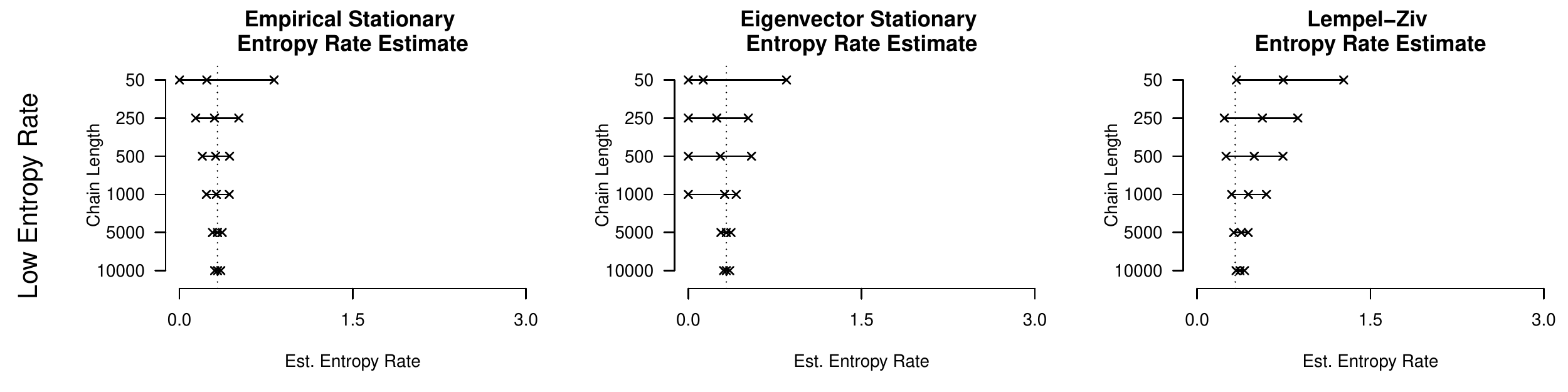}
\includegraphics[width=0.9\linewidth]{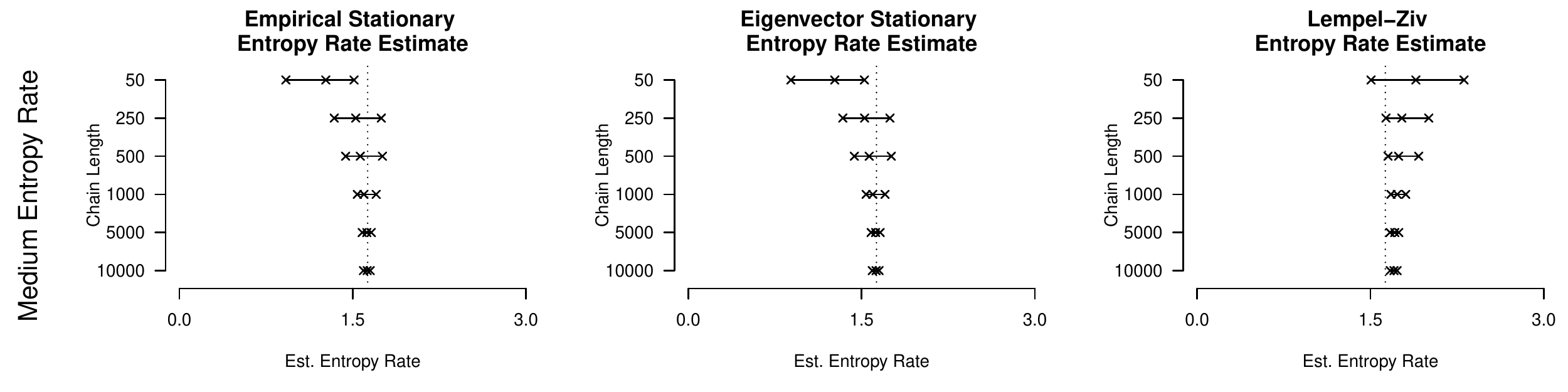}
\includegraphics[width=0.9\linewidth]{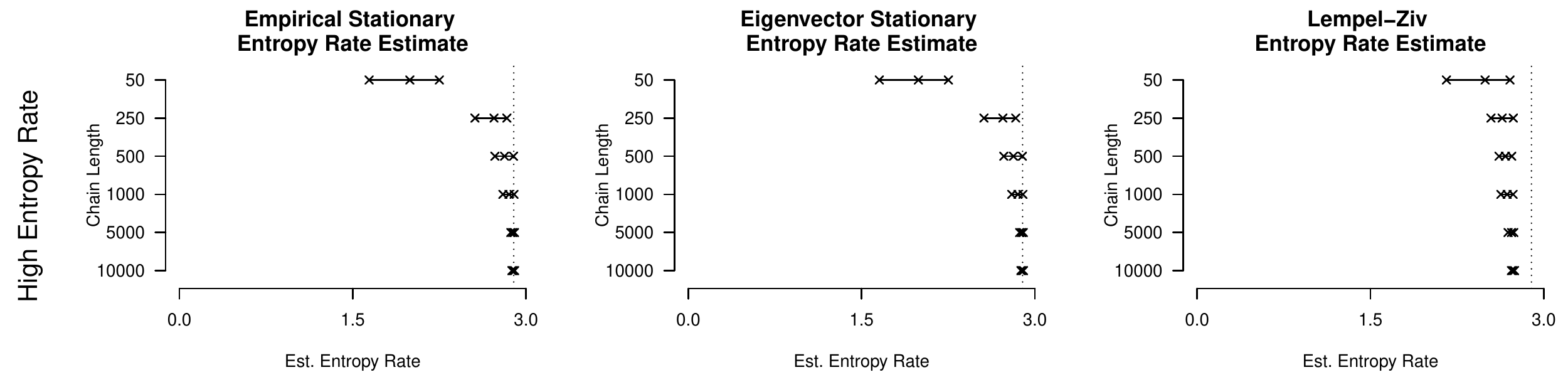}
\caption{Simulation results for 100 simulated Markov processes for each of the transition matrices shown in Figure \ref{fig:transitionmatrices}, ordered from top to bottom: low entropy rate, medium entropy rate, and high entropy rate.  Entropy rate estimates were obtained at subsequence lengths 50, 250, 500, 1000, 5000, 10000. Each column of the figure represents a different entropy rate estimation technique.}
\label{fig:montecarlo}
\end{figure}

Figure \ref{fig:montecarlo} provides the results from the simulation study and is organized as follows.  The top row of the figure contains results for the low entropy rate data generating model, followed by the second row which provides results for the medium entropy rate data generating model.  The final row gives results for the high entropy rate model. The first column of the figure demonstrates the performance of directly estimating the entropy rate with an empirical estimate of the transition matrix and of the stationary distribution, while the second column illustrates the performance of using an empirical estimate of the transition matrix and an eigendecomposition of the transpose of the observed transition matrix to estimate the stationary distribution of the process.  The last column provides the performance of the estimator based on the SWLZ algorithm. Each line within a subplot displays the spread of the entropy rate estimates from 100 realizations of the process. The marks on the line represent the lowest entropy rate, mean entropy rate, and highest entropy rate observed across these 100 realizations.  A primary result of the simulation study is that estimates from almost all realizations converge to a common value for long sequences, across all of the transition matrices.  

The first row of Figure \ref{fig:montecarlo} gives the performance of the three estimators in estimating the entropy rate of structured and predictable systems, i.e. low entropy rate systems.  We see that results are consistent with the true entropy rate when we empirically estimate the transition matrix and the stationary distribution from the observed sequence.  For this estimator, the average across the simulations is close to the true entropy rate even for relatively short sequence lengths of 250 and appears to be an extremely reliable estimate of the true entropy rate for sequences of length 5000 or greater.  When we estimate the stationary distribution by an eigendecomposition of $\bmath{ \hat P}^{T}$  and then estimate the entropy rate for low entropy rate systems, the second plot in the first row, we notice some performance issues.  In this case, we occasionally observe estimates of the entropy rate which are zero.  This is a result of the transition matrix that we have chosen for this example. From Figure \ref{fig:transitionmatrices}, there is a very high probability of a state being followed by the same state ($P_{ii}=0.95$). With shorter sequences it is therefore possible that we have not observed a transition out of one or more states when the sequence terminates. This creates a reducible transition matrix and the results for stationary distributions defined by eigenvalues of transition matrices are no longer valid, resulting in an entropy rate estimate of zero in our simulations.  For longer sequences this approach works well.  Additionally, we see the SWLZ estimates appear to be biased high for low entropy rate systems, but approaches the true entropy rate as sequence length increases.  Recall from Equation (\ref{eqn:lztheory}), that the theoretical results show that $\frac{\log_2(n)}{L_{n}(\mathcal{X})} \rightarrow_{P} H(\mathcal{X}), \mbox{ as } n \rightarrow \infty$ and that our estimator in Equation (\ref{eqn:lzestimator}) is $\hat H(\mathcal{X}) = \log_2n / \big(\frac{1}{n}\sum_{i=1}^n \Lambda_i(\mathcal{X})\big)$.  The numerator, $\log_{2}(n)$ appears because the underlying theory effectively assumes that there is a history of size $n$ when finding the length of the unique block $\Lambda_{i}(\mathcal{X})$. In truth though the algorithm only has at its disposal a history of size $i<n$.  Due to this limited history, the entropy rate is overestimated because the lengths of unique subsequences are shorter than expected for a history of size $n$.  As the sequence length increases, this becomes less of a concern and the estimates approach the true entropy rate.  

Now consider estimation of systems which are highly unpredictable, the last row of Figure \ref{fig:montecarlo}.  When the entropy rate is estimated using an empirical estimate of the transition matrix and stationary distribution, the mean of the estimates approach the true entropy rate from below as sequence length increases. This suggests that this method systematically underestimates the randomness of the system for short sequence lengths. Also, compared with low entropy rate systems, it takes more observations (i.e. $1000$ vs. $250$) for the mean of the estimates to converge to the true entropy rate. The same pattern holds when the stationary distribution is estimated by an eigendecomposition of the transpose of the empirical transition matrix, the second plot in this row.  The SWLZ estimator has difficulties estimating the entropy rate in high entropy rate systems that appear similar to the difficulties for low entropy rate systems, but in the opposite direction.  Instead of being biased high, the entropy rate estimate is biased below the truth.  Using a similar argument as was used for low entropy rate systems, this implies that on average the unique subsequence lengths obtained are longer than expected.  One suggested explanation for the bias is the fact that we can only find strings of integer length. Equation (\ref{eqn:lztheory}) implies that we should expect unique strings of mean length of approximately 3.44 for a sequence of length 1000 to achieve $H(X) = 2.8934$.   Finding unique strings of length 2 or 3 when using a history of 1000 observations is less likely than unique strings of length 4 or 5, and therefore we obtain longer than needed strings.  We note additional simulations (not shown) indicate that this bias is still present for sequences of length 50000.  

Intermediate entropy rate systems, the middle row of Figure \ref{fig:montecarlo}, have similar performance to the other rows of the figure.  For both methods of direct estimation, the results agree and approach the true rate from below.  Finally, it is apparent that the entropy rate estimates based on SWLZ are biased high in the intermediate entropy rate case, similar to the low entropy rate case. Estimates based upon the SWLZ estimator are approaching the true entropy rate, but we have not observed sequences long enough to assess true convergence.  

\subsection{Estimating the Standard Error via Stationary Bootstrap}\label{sec:bootstrapsims}

The previous section highlighted the performance of the methods in providing a point estimate of the entropy rate of the process.  In this section, we investigate the performance of the stationary bootstrap in estimating the standard error of the entropy rate estimate.  The simulation setup is the same as the previous section. The estimate of the entropy rate is used to select the parameter $p$ for the stationary bootstrap such that $p = \hat H(\mathcal{X}) / \log_{2}(j)$, where $j$ is the sequence length.  One hundred bootstrap replicates of each of the 100 simulated series were constructed as outlined in Section \ref{sec:bootstrap}.  The entropy rate was estimated on these 100 bootstrap samples and the standard error was calculated from these samples and recorded.  Thus we have 100 different bootstrap standard error estimates for simulation scenario.  We use the empirical standard error of the entropy rate estimates across the 100 simulations of Section \ref{sec:firstordersims} as a reference for assessing the performance of the bootstrap standard errors. 

\begin{figure}[h!]
\centering
\includegraphics[width=0.9\linewidth]{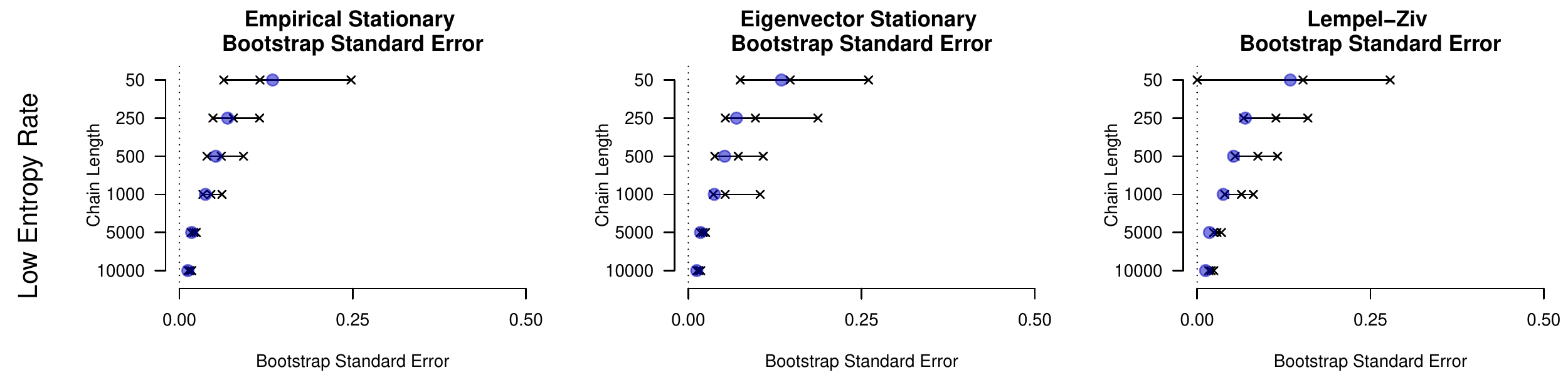}
\includegraphics[width=0.9\linewidth]{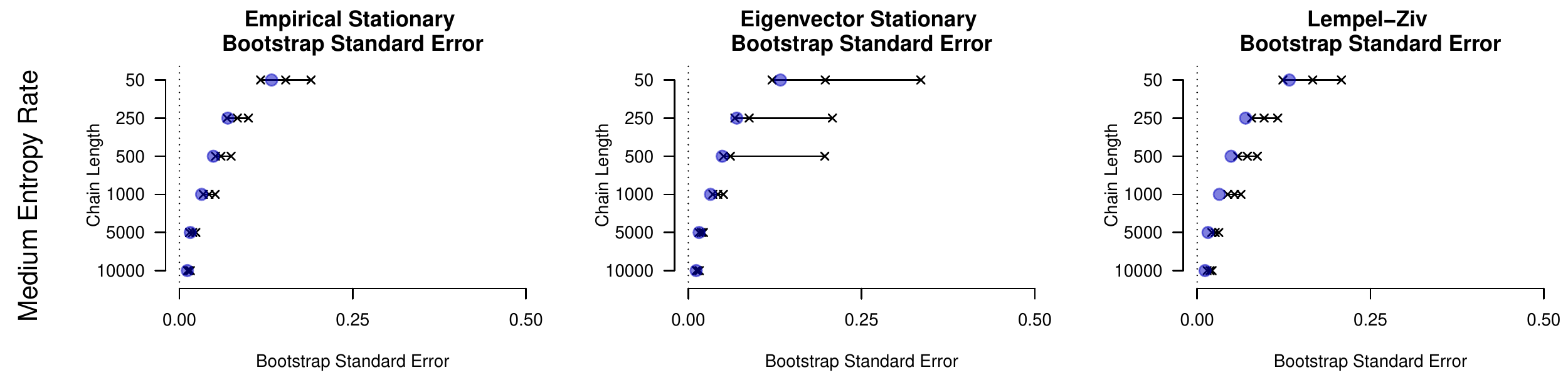}
\includegraphics[width=0.9\linewidth]{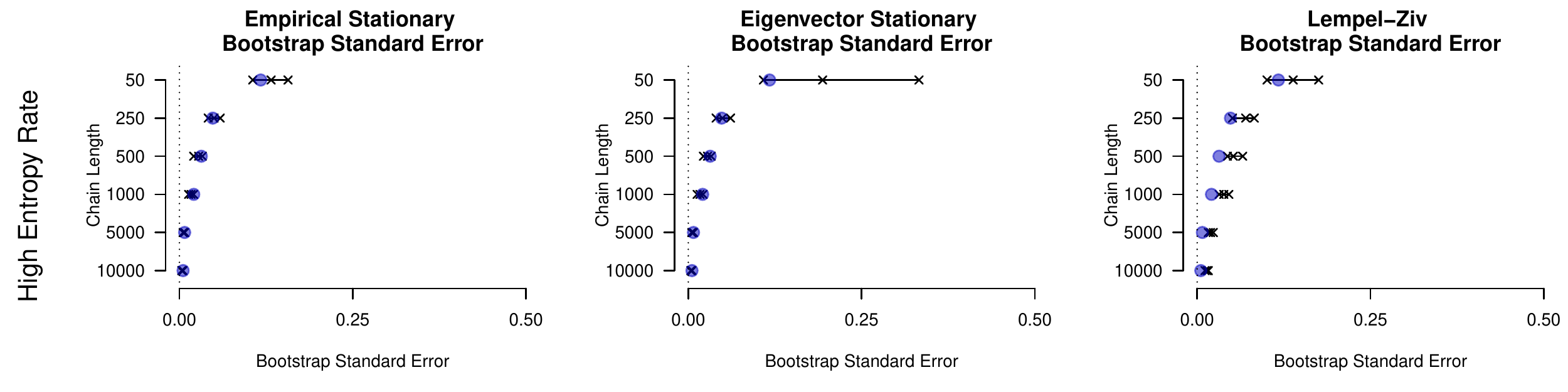}
\caption{Simulation results for using the stationary bootstrap to estimate the standard error of the entropy rate estimate. Ordered from top to bottom: low entropy rate, medium entropy rate, and high entropy rate.  Bootstrap standard errors were obtained for each of 100 simulated Markov processes at subsequence lengths 50, 250, 500, 1000, 5000, 10000. Each column of the figure represents a different entropy rate estimation technique.  In each simulation the entropy was estimated and the parameter $p$ for the stationary bootstrap was selected as described in Section \ref{sec:bootstrap}. The blue dots represent the empirical standard error from Section \ref{sec:firstordersims} using the empirical estimation of the stationary distribution.}
\label{fig:bootstraperrors}
\end{figure}

Figure \ref{fig:bootstraperrors} provides the results of the standard error estimates and is organized similarly to Figure \ref{fig:montecarlo}.  The first result from these figures is that the empirical standard error (the blue dots) agrees with the distribution of bootstrap standard errors (represented by the x's and lines). We see from the figure that in most cases the standard errors obtained from the bootstrap procedure are greater than the empirical standard error and therefore provides a conservative estimate of the standard error.  

\subsection{Estimation When $m$ is Misspecified}\label{sec:secondordersims}

Section \ref{sec:firstordersims} suggests that direct estimation of the entropy rate works extremely well when the order of the Markov chain is specified correctly and the SWLZ estimator is less efficient.  Now we consider estimating the entropy rate when the true data generating process is a second-order Markov chain, but we do not necessarily assume this is known.  In this simulation scenario, we compare the performance of four estimators of the entropy rate: direct estimation assuming (incorrectly) a first-order process, direct estimation assuming (correctly) a second-order process, direct estimation assuming a third-order process, and estimation using the SWLZ compression algorithm.  This scenario highlights the advantage of the SWLZ method for estimating the entropy rate when the order of the process is unknown.  

\begin{figure}
\centering
\includegraphics[width=0.8\linewidth]{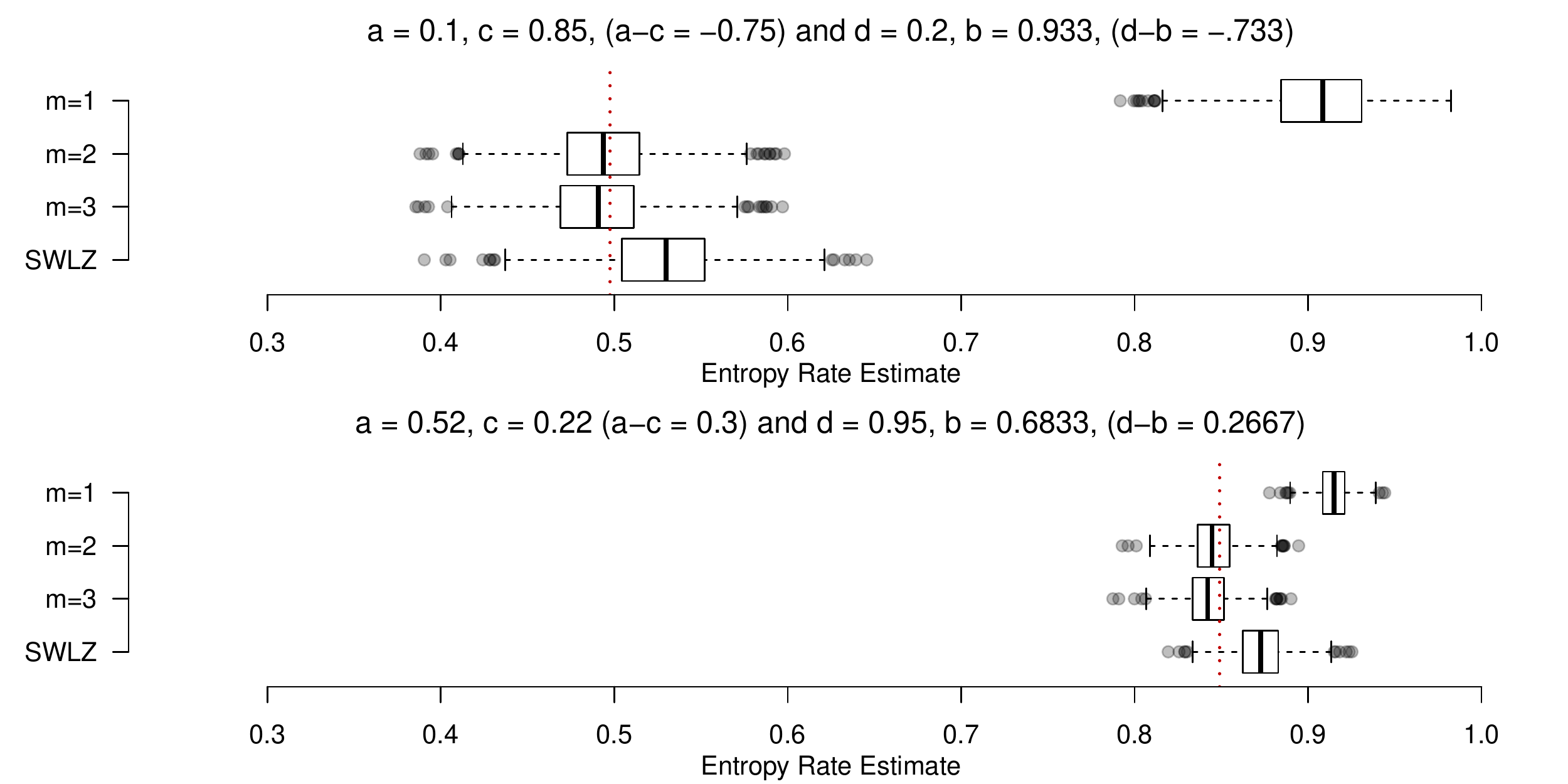}
\caption{Comparing the estimated entropy rate of a two-state Markov chain when the order is specified as $m=1,2,3$ and without specification in the SWLZ method, the true entropy rate is the vertically dotted line in each figure.  The upper figure contains an extreme case when the magnitudes of $\phi$ and $\gamma$ are large.  The lower figure provides an example when the magnitudes of $\phi$ and $\gamma$ are approximately 0.3.  We see that misspecification results in a bias in estimating the entropy when we choose an $m$ which is too small.}
\label{fig:secondordersims}
\end{figure}

Assume that the true process is a second-order time-homogeneous Markov process on two states which we label $A,B$. Then for $\alpha_i, \alpha_j, \alpha_k \in \mathcal{A}=\{A,B\}$, the transition probabilities can be written,
\begin{eqnarray*}
	& &\mbox{Pr}(X_t = \alpha_k | X_{t-1} = \alpha_j, X_{t-2} = \alpha_i) \\
	&=& \mbox{Pr}(X_t = \alpha_k,  X_{t-1} = \alpha_j | X_{t-1} = \alpha_j, X_{t-2} = \alpha_i) \\
	&=& P(\alpha_i\alpha_j, \alpha_j\alpha_k) =P_{ij, jk}.
\end{eqnarray*}
The transition matrix of this second-order system can be organized as follows,
\[
	\bmath{P}_{(2)}
			= \left( \begin{array}{cccc}
				P(AA, AA) & P(AA, AB) & 0 & 0 \\
				0 & 0 & P(AB, BA) & P(AB, BB) \\
				P(BA, AA) & P(BA, AB) & 0 & 0 \\
				0 & 0 & P(BB, BA) & P(BB, BB) \\
				\end{array} \right)
\]
where transitions that are not possible, e.g. from $AA$ to $BA$, are assigned probability zero.  Here the notation, i.e.  $P_{(2)}$, denotes that this is a transition matrix for a second-order Markov chain and will be utilized for clarity.   Similarly, for this process, the row vector representing the stationary distribution is,
\[
\pi_{(2)} = (\pi_2(AA), \pi_2(AB), \pi_2(BA), \pi_2(BB)).
\]
Provided the transition matrix specified is irreducible, the stationary distribution exists and represents the joint distribution of consecutive observations, $X_{t-1}$ and $X_{t}$.  Now, assume the following parameterization for $\bmath{P}_{(2)}$,
\[
	\bmath{P}_{(2)} = \left( \begin{array}{cccc}
				1-a & a & 0 & 0 \\
				0 & 0 & b & 1-b \\
				1-c & c & 0 & 0 \\
				0 & 0 & d & 1-d \\
				\end{array} \right).
\]
If $a=c$ and $b=d$ then the process behavior is indistinguishable from that of a first-order Markov process.  To see this note that the transition probabilities from $AA$ and $BA$ are identical so that only the most recent state matters. Thus differences between $a$ and $c$ (or $b$ and $d$) control how much the process depends on the observation which occurred two time steps ago.  Using this parameterization, we simulate 1000 second-order Markov chains, each sequence consisting of 1000 observations, and estimate the entropy rate of the process directly assuming orders $m=1,2,3$ and additionally using the SWLZ estimator which makes no assumption on the order of the process.  We consider two cases which we label as Case I and Case II: (I) $a = 0.1, c = 0.85, d = 0.2, b = 0.933$ and (II) $a = 0.52, c = 0.22, d = 0.95, b = 0.6833$.  The first case highlights an extreme example when the difference between $a$ and $c$ (or $b$ and $d$) is large and the Markov process depends almost exclusively on the observation which occurred two time steps ago.  The difference between the parameters $a$ and $c$ (or $b$ and $d$) is smaller in the second case.  Simulation results are in provided in Figure \ref{fig:secondordersims}


The simulations suggest that when the order of the Markov chain specified is too low (i.e., $m=1$), the estimate obtained will be biased above the true entropy rate (unless $a=c$ and $b=d$).  In contrast when we provide the correct order ($m=2$), or an order which is too large $(m=3)$, we obtain unbiased results.   The SWLZ estimator accurately estimates the entropy rate of the process without requiring that the order be specified, but is biased in short sequences (as discussed in Section \ref{sec:firstordersims}).  

These results also suggest that the bias from an $m$ which is too low, increases as the difference between the values of $a$ and $c$ (or $b$ and $d$) increases.  
This relationship can actually be quantified.  We can derive the first-order dependence structure that would be observed from monitoring transitions of a second-order process for a long time.  This depends on $a,b,c,d$ through the stationary distribution of the second-order process defined by $\bmath{P}_{(2)}$.  An eigendecomposition of $\bmath{P}^{T}_{(2)}$ yields, $\pi_{(2)} = \psi^{-1} \left( d(1-c),da, da,a(1-b) \right)$ where $\psi=a(1-b) + 2da + d(1-c)$.  It can then be shown that the first-order dependence structure of the second-order process is,
\[
	\bmath{P}_{(1)} = \left( \begin{array}{cc}				
			\frac{1-c}{(1-c) + a} & \frac{a}{(1-c) + a} \\
			\frac{d}{d + (1-b)} & \frac{1-b}{d + (1-b)}
	\end{array} \right) 
\]

\begin{figure}
\centering
\includegraphics[width=0.6\linewidth]{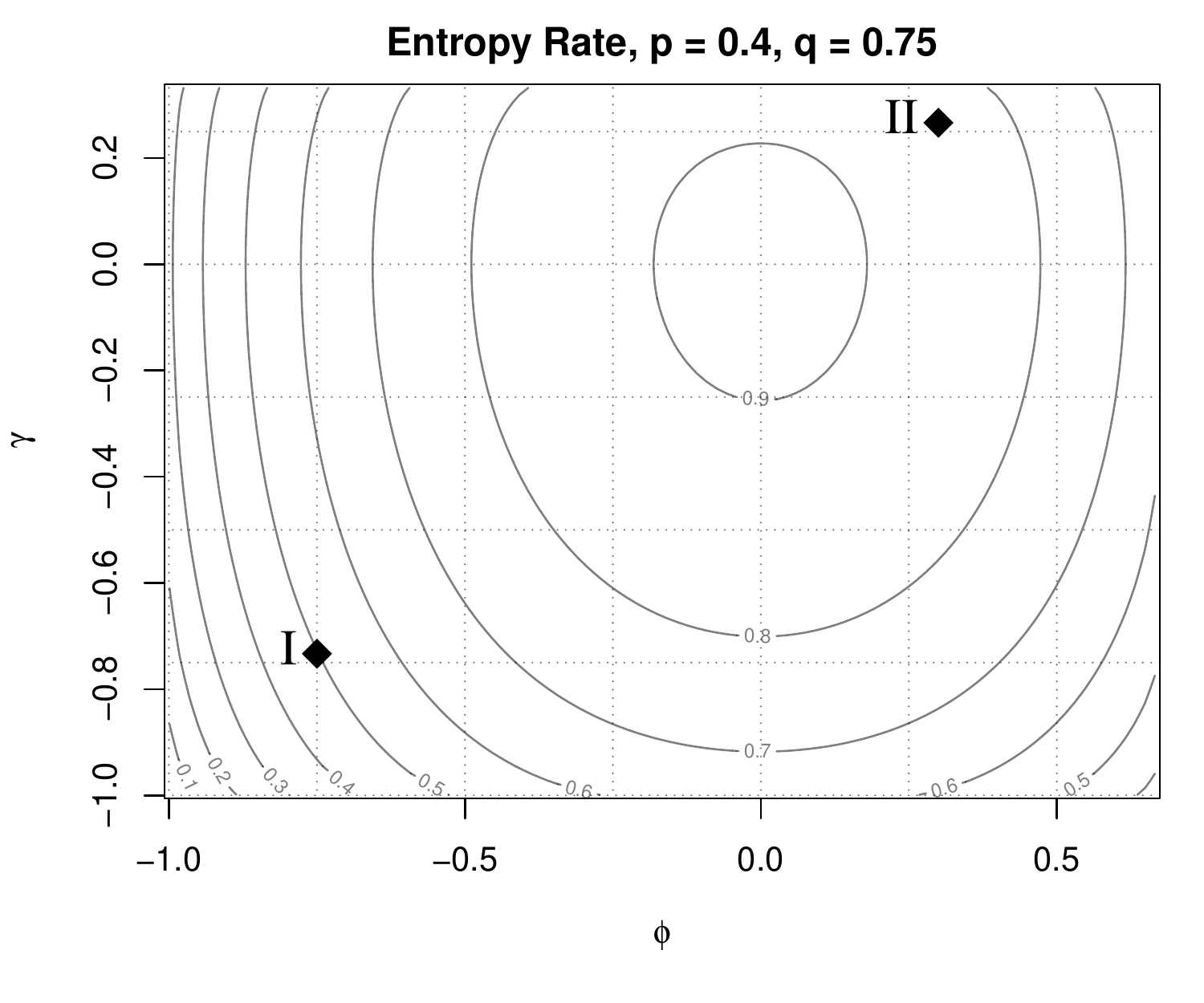}
\caption{Second-order entropy rate estimates for $p=0.4$ and $q=0.75$ as well as a grid of points for valid values of $\phi$ and $\gamma$. The cases used in out simulations are identified on the figure. We find that for even reasonably large values of the tuning parameters, $|\phi| \approx 0.5$ and $-0.75 < \gamma < 0.3$, the difference between the first order entropy estimate and the second order entropy estimate will be moderate (as in our Case II).}
\label{fig:twoordertruth}
\end{figure}

If we misspecify the order of the process, then we only observe the entries of the first-order transition matrix.  Then if we write, 
\[
	\bmath{P}^{*}_{(1)} = \left( \begin{array}{cc}
				1-p & p \\
				q & 1-q \\
				\end{array} \right).
\]
and define, $\phi = a - c$ and $\gamma = d-b$, we can show that
\[
\begin{array}{cc}
a = p(1+\phi)  & (1-c) = (1-p)(1 + \phi) \\
d = q(1 + \gamma) & (1-b) = (1-q)(1+\gamma)
\end{array}.
\]
which implies the following restrictions on $\phi$ and $\gamma$:  $-1 \le \phi \le \min(\frac{p}{1-p}, \frac{1-p}{p})$ and $-1 \le \gamma \le \min(\frac{q}{1-q}, \frac{1-q}{q})$.  Now we can rewrite $\bmath{P}_{(2)}$ in terms of the observed entries of the first-order transition matrix, $p, q$ and the parameters $\phi, \gamma$, 
\begin{equation}
	\bmath{P}_{(2)} = \left( \begin{array}{cccc}
				(1 + \phi) \left(\frac{1}{1 + \phi} - p\right)  & p(1 + \phi) & 0 & 0 \\
				0 & 0 & (1 + \gamma)\left(q - \frac{\gamma}{1 + \gamma}\right) & (1-q)(1+\gamma) \\
				(1-p)(1+\phi) & (1 + \phi)\left(p - \frac{\phi}{1 + \phi}\right) & 0 & 0 \\
				0 & 0 & q(1 + \gamma) & (1 + \gamma)\left(\frac{1}{1 + \gamma} - q\right) \\
				\end{array} \right) 
\label{eqn:secondorderparameter}
\end{equation}
Both second-order simulations were constructed to have the same first-order behavior with $p=0.4$ and $q=0.75$. Equation (\ref{eqn:secondorderparameter}) allows an analytical calculation of the entropy rate for any combination of $\phi$ and $\gamma$ for fixed $p$ and $q$.  The contour lines of Figure \ref{fig:twoordertruth} are the entropy rates of the second order process when $p=0.4$ and $q=0.75$ for a grid of $\phi$ and $\gamma$.   The max of this plot occurs when $\phi=\gamma=0$ (a first-order process) and at this point $H(\mathcal{X}) = 0.915$.  Additionally marked in the figure are the values of $\phi$ and $\gamma$ for the simulations of this section.  We see that for a large portion of the figure, the second order process will have an entropy rate greater than $0.8$ and therefore the bias from mistakenly using a first-order model will be small (as seen in the example given in the bottom panel of Figure \ref{fig:secondordersims}).  

\section{Measuring Predictability of Maternal Care in Rodents}

We now provide an application of entropy rate estimation using data that was introduced in  \cite{molet2016fragmentation}.  The study addressed the impact of fragmented and unpredictable behavior of rodent mothers on the emotional and cognitive outcomes of their offspring. On postnatal day 2 (P2), rodent pups were randomly assigned to two types of rearing environments for 8 days; a normal environment and an impoverished environment (limited bedding and nesting materials).  The dams (mothers) were observed twice per day for 50 minutes for eight days and sequences of their behavior were recorded.  Behavior was described using a set of $\kappa = 7$  distinct actions: licking/grooming pups, carrying pups, nursing, nest building, off pups, eating, or self-grooming.  On the tenth day (P10), the rodent pups and mothers were returned to normal environmental conditions.  

The limited bedding and nesting environment led to more erratic maternal behavior and a goal was to quantify the impact of this treatment (i.e. the unpredictability of maternal care due to an adverse environment).  Entropy rate was chosen as the measure which could most succinctly summarize the behavior of each dam, rather than focusing on any specific action that the rodents perform.  In \cite{molet2016fragmentation}, the sequence of observations for each rodent was treated as a stationary first-order Markov chain.  The assumption of stationarity allowed the concatenation of the sequences from each 50 minute window together as one long Markov process.  Because the analysis was focused on the predictability of the \textit{patterns} of actions and not the duration of actions, the sequence was treated as a discrete-time Markov chain focusing only on transitions between different maternal care behaviors.  Under these assumptions, estimates of entropy rate were calculated for each rodent. Note that there were a total of seven possible actions which corresponds to a maximum possible entropy of $H_{\max}(\mathcal{X}) = \log_2(7) = 2.807$.  

\begin{figure}
\centering
\includegraphics[width=0.4\linewidth]{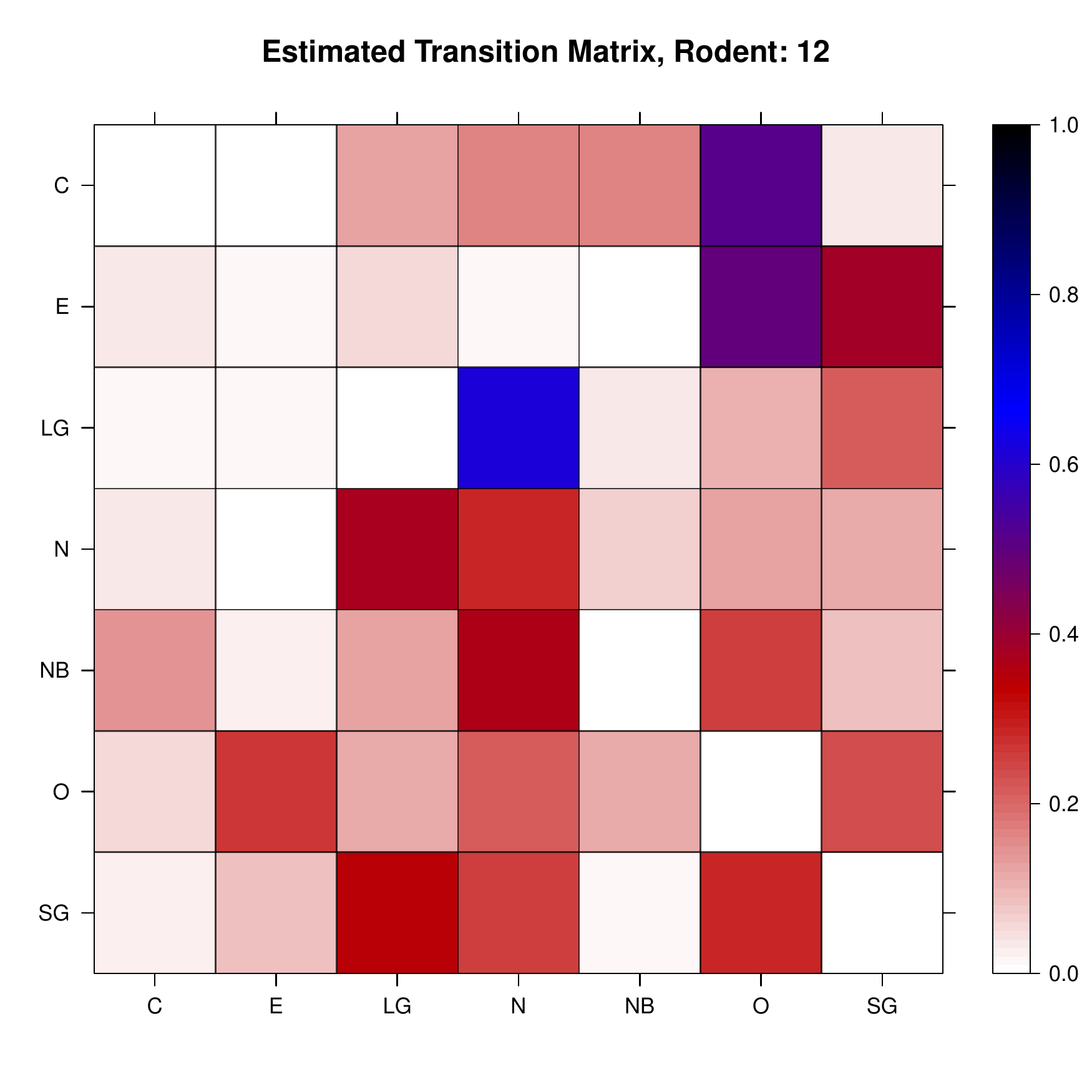}\includegraphics[width=0.4\linewidth]{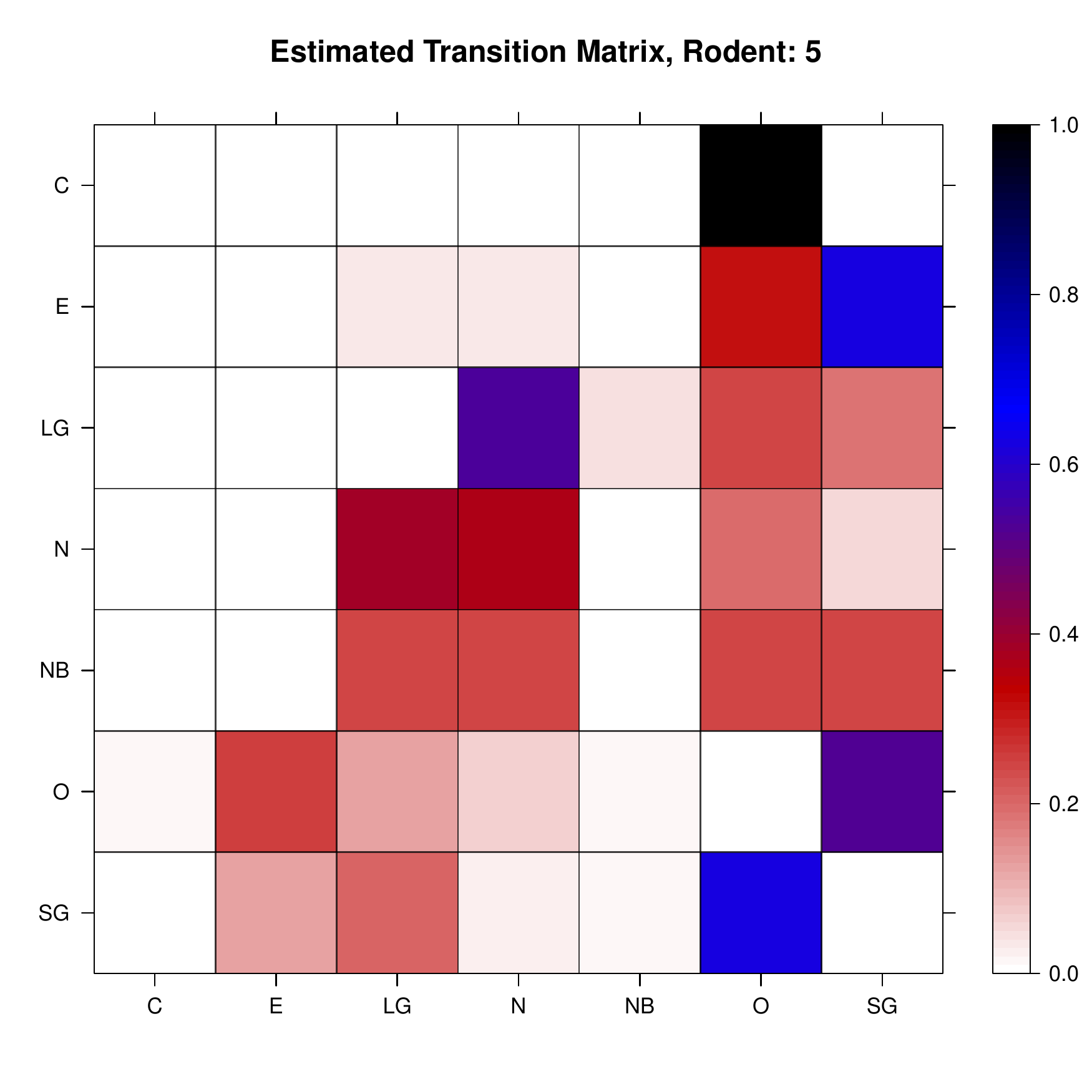}
\caption{Examples of observed empirical transition matrices from the behavioral application.  Left: example limited bedding \& nesting rodent, Right: example control rodent}
\label{fig:treatmenttransition}
\end{figure}

In this paper, we recreate these analyses using the estimators that were outlined in the previous sections.  We obtain results that are consistent with those obtained in the original paper.  Figure  \ref{fig:treatmenttransition} provides an empirical transition matrix from the limited bedding, and nesting group and one from the control group respectively, as an example of the data in the study.  

To summarize the level of predictability, Table \ref{tab:rodent} provides entropy rate estimates based upon treating the rodent behavior as a first-order Markov chain, a second-order chain, and by using the sliding-window Lempel-Ziv estimator not assuming any specific order for the behavior.  We see that the minimum number of transitions performed by any rodent was approximately 300 which, by our simulation study, ensures that the estimates will be adequate for the task at hand.  The series is not long enough to consider a third-order process, $7^3 = 343 > 300$. \cite{molet2016fragmentation} performed a $t$-test on the difference in means between the entropy rates of the two groups which showed that experimentally introducing a stressed environment for rodent mothers leads to less predictable maternal care.

\begin{table}
\centering
\caption{Rodent Data Estimates}
\begin{tabular}{c|c|c|c|cc|cc}
 \hline
ID & Group & $N$ & $\hat H_{SWLZ}(\mathcal{X})$ & $\hat H_{emp}^{m=1}(\mathcal{X})$ & $\hat H_{eig}^{m=1}(\mathcal{X})$ &  $\hat H_{emp}^{m=2}(\mathcal{X})$ & $\hat H_{eig}^{m=2}(\mathcal{X})$ \\ 
  \hline
  \hline
  3 & LBN & 511 & 1.6956 & 1.8837 & 1.8831  & 1.5069 & 1.5060 \\ 
  4 & LBN & 404 & 1.6285 & 1.8015 & 1.8009 & 1.3307 & 1.3298 \\ 
  7 & LBN & 688 & 1.6797 & 1.8774 & 1.8770 & 1.4988 & 1.4986 \\ 
  8 & LBN & 699 & 1.6807 & 1.8403 & 1.8399 & 1.5168 & 1.5150 \\ 
  11 & LBN & 632 & 1.7916 & 1.9342 & 1.9338 & 1.6192 &  1.6171 \\ 
  12 & LBN & 886 & 1.8526 & 2.0515 &  2.0511 & 1.7462 & 1.7462 \\ 
\hline
\multicolumn{3}{r}{Group Mean} & 1.7215&1.8981&1.8976&1.5364&1.5354 \\
\hline
\\
 \hline
ID & Group & $N$ & $\hat H_{SWLZ}(\mathcal{X})$ & $\hat H_{emp}^{m=1}(\mathcal{X})$ & $\hat H_{eig}^{m=1}(\mathcal{X})$ &  $\hat H_{emp}^{m=2}(\mathcal{X})$ & $\hat H_{eig}^{m=2}(\mathcal{X})$ \\ 
\hline
1 & Control & 433 & 1.5483 & 1.7393 & 1.7395 & 1.3632 & 1.3627 \\ 
  2 & Control & 340 & 1.5107 & 1.5322 & 1.5319 & 1.2044 &  1.2042 \\ 
  5 & Control & 290 & 1.5727 & 1.6256 & 1.6242 & 1.2621 &  1.2596 \\ 
  6 & Control & 378 & 1.6571 & 1.7427 & 1.7417 & 1.3900 & 1.3878 \\ 
  9 & Control & 300 & 1.7552 & 1.8164 & 1.8155 &  1.3637 &1.3536 \\ 
  10 & Control & 403 & 1.7864 & 1.8590 & 1.8591  & 1.4391 & 1.4326 \\ 
   \hline
\multicolumn{3}{r}{Group Mean} &  1.6384&1.7192&1.7187&1.3371&1.3334 \\
\hline
\\
\hline
\multicolumn{8}{l}{$t$-test Results - Difference in Means with Equal Variance}  \\
\hline
\multicolumn{3}{r}{Test Statistic} &  -1.4425 & -2.9308 & -2.9287 & -2.986 & -3.0428 \\
\hline
\end{tabular}
\label{tab:rodent}
\bigskip
\end{table}


Table \ref{tab:rodent} shows that in this setting, we obtain results consistent with those of \cite{molet2016fragmentation} for first-order Markov chains.  Additionally, we see that SWLZ estimation provides estimates which fall between the estimates obtained using first-order and second-order assumptions.  This may imply that the behavior of the rodents is more complex than the simplifying assumption of first-order behavior and that the rodent behavior is better represented by a second-order process.  The correlation between the first-order and second-order entropy rates based on the empirically estimated transition matrix and stationary distribution is approximately 0.937.

\section{Discussion}

Behavior can be modeled as a sequence of actions performed by an individual over a given time period.  The entropy rate of this sequence is a summary measure that describes the degree to which we can predict actions of the process.  In this paper we presented three different entropy rate estimators and assessed their performance as a function of the length of observed sequences.  Additionally, we presented the stationary bootstrap as an approach for obtaining standard error estimates, provided a method for choosing the parameter of the stationary bootstrap, and assessed the performance of the stationary bootstrap.    

Estimating the entropy rate by direct estimation of the transition matrix and stationary distribution of a Markov process achieves asymptotically unbiased estimates of the true entropy rate when the order of the Markov process is known. This approach provides several summary measures of behavior.  We obtain a description of the probability distribution governing transitions between actions, the long term expected proportion of occurrences that each action is performed, and a measure of the predictability of actions.  There are two major concerns associated with using direct methods to estimate the entropy rate. One concern is that we must observe $n \gg \kappa^m$ observations to achieve precise estimates of the transition matrix for an $m^{th}$ order Markov process.  It is often challenging to observe a sequence of behaviors long enough to achieve the required precision when $m$ is 3 or more.  A more serious concern is the requirement that the order of the Markov process is assumed known.  There is always the possibility that the true data generating process does not match the assumed order.  If the true order is larger than the assumed order, then one will get misleading results.  
%

The sliding-window Lempel-Ziv (SWLZ) estimator avoids the assumption of a specific order for the Markov chain.  It is based on a data compression algorithm that only assumes that the stochastic process is both stationary and ergodic.  It appears from the simulation study that estimates based upon this strategy can be slightly biased, particularly in short sequences.  Section \ref{sec:secondordersims} demonstrates the advantage of the approach in that the SWLZ method accurately estimates the true entropy rate for higher-order processes without requiring that the order be specified in advance.  

The two estimation techniques may be used in concert to fully understand the predictability of an observed stochastic system. They both provide valuable information regarding predictability.  The SWLZ method can be used to provide an initial estimate of the entropy rate which is assured of being close to the true entropy rate of the system and therefore may be used as a guidepost for choosing the order $m$ of the process.  Once we have an estimate of the entropy rate, we can directly estimate the entropy rate for $m=1,2,\dots$ provided we have observed enough data and until the entropy rate estimate is approximately equal to that obtained from the SWLZ technique.  This would allow us to further understand the process through its transition matrix and  stationary distribution.  

Each of the methods rely on the assumption that the process is stationary. This critical assumption is easy to violate.  Biological systems, like those described in our application, may not be stationary since the predictability of the individual may be context dependent.  For example, an individual may be predictable when in a familiar environment, but the individual may be unpredictable if moved into a new environment.  This implies that when we intend to use entropy rate to define behavioral characteristics, we should restrict our observations to specific windows of time that are long enough to estimate the entropy rate, but short enough for the assumption of stationarity to be plausible.

\backmatter


\section*{Acknowledgements}

The first author would like to thank Dr. Alexander Vandenberg-Rodes for his help in developing a deeper understanding of Markov processes. Research reported in this article was supported by NIMH Silvio O. Conte Centers of the National Institutes of Health under award number P50 MH 096889. The content is solely the responsibility of the authors and does not necessarily represent the official views of the National Institutes of Health.\vspace*{-8pt}


%


%
\bibliographystyle{draftstyle}
\bibliography{estimatingentropyrate}{}
%
%
%
%

%
%
%

\label{lastpage}

\end{document}